1

# Deep Learning Framework for Multi-Round Service Bundle Recommendation in Iterative Mashup Development

Yutao Ma, Xiao Geng, Jian Wang, Keqing He, and Dionysis Athanasopoulos

*Abstract*—Recent years have witnessed the rapid development of service-oriented computing technologies. The boom of Web services increases software developers' selection burden in developing new service-based systems such as mashups. Timely recommending appropriate component services for developers to build new mashups has become a fundamental problem in service-oriented software engineering. Existing service recommendation approaches are mainly designed for mashup development in the single-round scenario. It is hard for them to effectively update recommendation results according to developers' requirements and behaviors (e.g., instant service selection). To address this issue, we propose a service bundle recommendation framework based on deep learning, DLISR, which aims to capture the interactions among the target mashup to build, selected (component) services, and the following service to recommend. Moreover, an attention mechanism is employed in DLISR to weigh selected services when recommending a candidate service. We also design two separate models for learning interactions from the perspectives of content and invocation history, respectively, and a hybrid model called HISR. Experiments on a real-world dataset indicate that HISR can outperform several state-of-the-art service recommendation methods to develop new mashups iteratively.

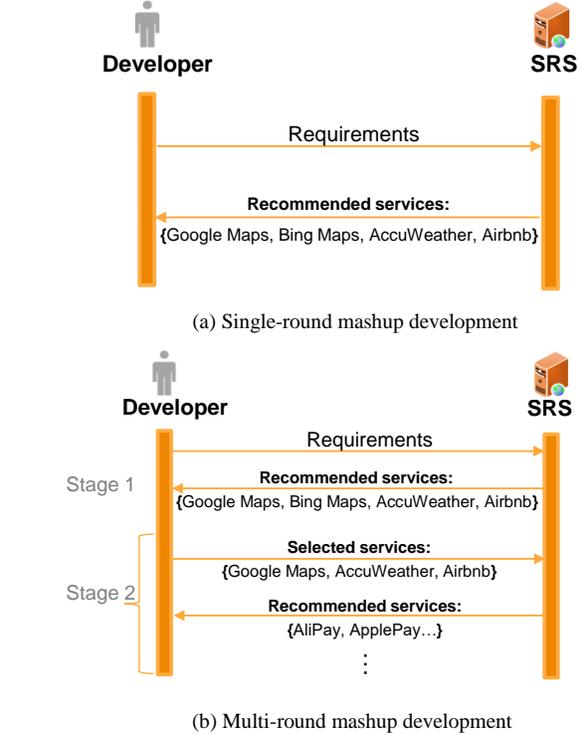

(a) Single-round mashup development

(b) Multi-round mashup development

Fig. 1. Two different scenarios for mashup development.

## I. INTRODUCTION

With the maturity of the service-oriented computing paradigm, service-oriented software development has increasingly become popular. A large number of Web services have been released on the Internet. By integrating these existing Web services, software developers can build their mashups (i.e., Web applications that can provide certain functionalities by composing one or more services) more efficiently. Until now, this methodology has shown ample ability to reduce the cost of system development and increase the quality of service-based software [1]. However, the rapid growth of Web services available in Web application programming interface (API) directories such as ProgrammableWeb[1] raises challenges in selecting suitable services for mashup development. Service recommendation has emerged as a critical technology that recommends appropriate component services for developers in the development process of mashups.

Many service recommendation methods have been proposed in the past decade. They are roughly classified into three main categories: content-based, collaborative filtering (CF)-based, and hybrid approaches, according to the information used in the recommendation process. The content-based approaches [2]-[4] make recommendations based on the textual similarities between service descriptions and user requirements. The CF-based approaches [5]-[8] leverage the historical experience of similar mashups/services to generate a recommendation list. By integrating content-based and CF-based approaches, the hybrid approaches [9]-[15] consider explicitly-specified requirements, implicit invocation preferences, and other information of service usages, such as co-invocation and popularity, to make recommendations.

The existing service recommendation approaches are mainly applicable for mashup development in the single-round scenario, i.e., they recommend a package of services (also called a service bundle) for a new mashup for one time. We take

Y. Ma, X. Geng, J. Wang, and K. He are with the School of Computer Science, Wuhan University, China. D. Athanasopoulos is with the School of Electronics, Electrical Engineering and Computer Science, Queen's University Belfast, UK.

E-mail: {ytma, xiaogeng5515, jianwang, hekeqing}@whu.edu.cn, d.athanasopoulos@qub.ac.uk.

[1] https://www.programmableweb.com/

developing a trip preparation mashup as an example. Assume that a developer's initial requirements are as follows: *A mashup can help tourists prepare for their trips. It can provide users with information about weather and tourist attractions, help them design travel routes, and book hotels online.* A service recommendation system (SRS) provides suggestions about component services in developing this new mashup. According to the initial requirements, in the single-round scenario (see Fig. 1(a)), the SRS might generate a recommendation list that contains Google Maps, Microsoft Bing Maps, AccuWeather, and Airbnb. However, the recommendation list may lack an electronic payment service for users to prepay online after booking a hotel. Due to the homogenization effect of services, it is also reluctant to recommend similar services with the same functionality, such as Google Maps and Microsoft Bing Maps.

Modern software development is generally viewed as an iterative and incremental process [16], and mashup development is no exception. This process requires recommending appropriate component services after receiving developer feedback in the multi-round scenario. As shown in Fig. 1(b), the iterative mashup development will generate a new recommendation list after receiving the developer's feedback on selected services. As a result, those services related closely to the selected services will have a higher priority in the new recommendation list. For example, the SRS is more likely to recommend an online payment service due to its high relevance with Airbnb, a selected hotel booking service. Moreover, since the developer has chosen Google Maps from the recommendation list, the SRS will respond to this feedback and most likely filter out other similar map APIs, e.g., Bing Maps, in subsequent recommendations. Then, in the second-round recommendation, the SRS may prioritize payment services like ApplePay and Alipay while lowering other map services' priorities.

A typical multi-round service bundle recommendation process consists of two stages. In the first stage, the SRS analyzes keyword-based developer requirements. It returns a list of candidate services, from which the target developer selects one or more candidate services. The recommendation process will enter the second stage if the result does not fully satisfy the requirements. The SRS continues to generate a new service list according to the requirements and existing component services that have been selected. This step will repeat until the requirements are satisfied.

According to developer feedback, the multi-round service bundle recommendation can potentially recommend appropriate component services to meet developer requirements. However, such an iterative scenario of mashup development presents some challenges to SRSs.

(1) An SRS deals with multiple interactions among a mashup, selected services, and the next service to recommend in the mashup development process. The candidate service must meet specific requirements of the mashup, and it should be as much as possible interoperable and interface-compatible [17] with selected services. Also, the interactions between the target mashup and its selected services may affect the selection of candidate services. User-based CF and content-based methods only consider the relationships between mashups and candidate services while ignoring the critical role of selected services in the recommendation process. Therefore, it is challenging for them to utilize developer feedback or update recommendation lists in the multi-round scenario.

(2) After a developer selects one or more component services, an SRS needs to make a prompt response to user feedback each time. It should accurately and timely capture and analyze developer requirements for the next round and generate follow-up recommendations. Supporting multi-round recommendations in a unified manner is another challenge for SRSs. Model-based CF methods, such as matrix factorization (MF) [18] and neural CF [19], have a minimal ability to model new mashups without component services. Even if developers have selected component services, it is hard to timely update recommendation models to obtain the latest representations of mashups.

*Given keyword-based developer requirements and selected component services, how does an SRS recommend appropriate services to developers in the mashup development process?* We propose a unified recommendation framework based on deep learning to address this issue. The framework aims to learn how a mashup, selected component services, and a candidate service interact and how their interactions affect a developer's selection of candidate services. It takes various types of information as input, e.g., content and historical invocation, to obtain feature representations of mashups, selected component services, and candidate services. Because each selected service has a different impact on selecting a candidate service, the framework also adopts an attention mechanism to weigh such impacts automatically. Besides, the framework can be easily applied to the single-round scenario by deactivating or masking its component designed for selected services.

In brief, the main contributions of this work are three-fold.

(1) Unlike the single-round scenario in services computing, we define and discuss a multi-round service recommendation scenario in the case of newly-selected component services for iterative mashup development with service bundles, which has not been sufficiently explored in the existing literature.

(2) We propose a service recommendation framework for the scenario mentioned above, called DLISR (short for deep learning and interaction-based service recommendation). This framework can gradually recommend appropriate component services to satisfy developer requirements by leveraging the information of selected services.

(3) Following the DLISR framework, we implement two separate models, which learn specific interactions using the content information of mashups and services and the historical invocations between them, respectively, and a hybrid model. Moreover, the experimental results on a real-world dataset indicate that the hybrid model outperforms other competitive baseline approaches in the multi-round scenario.

The rest of this paper is organized as follows. Section II introduces the related work. Section III presents the definition of the problems investigated in this study. Section IV details the proposed DLISR framework. Then, Section V describes three models that implement the proposed framework, and Section

VI presents the experimental results and analysis. Finally, Section VII summarizes this paper and puts forward our future work.

## II. RELATED WORK

### A. Single-round service recommendation

In the single-round scenario of mashup development, a service recommendation approach generates only a list of candidate services for the target mashup at a time. The existing approaches can be divided into three categories: content-based, CF-based, and hybrid.

The content-based approach predicts a service's rating over a mashup according to the similarity between service description and keyword-based developer requirements. WS-Finder [2], a keyword-based framework, applied the Earth Mover's Distance in multimedia databases to many-to-many partial matching between developer requirements and service attributes. Al-Hassan *et al.* [3] used domain ontologies to enrich the semantics of content information, and they measured the semantic similarity by logical reasoning. Shi *et al.* [4] utilized service labels to retrieve and highlight function-related words in service descriptions with the attention mechanism. They proposed a deep structured semantic model to measure the matching between a mashup and a service in functionality.

The CF-based approach mines implicit user preferences and generalizes patterns of similar users or items. For example, Samanta *et al.* [5] proposed an MF-based method for analyzing historical invocations between services and mashups. They took the inner product of latent factors as a critical factor in determining interaction probability. By applying the graph embedding technology to a user co-tag network and a social network, Wu *et al.* [6] obtained two different representations for user preference and social relationship. They integrated the two embeddings into a user-based CF recommendation model. Zou *et al.* [7] incorporated user-based and service-based CF in a reinforced CF framework that can remove services/users dissimilar to the target service/user when predicting the quality of service (QoS) values. Liang *et al.* [8] used a heterogeneous information network (HIN) to represent various entities, such as description, tag, and provider of mashups and services. They measured the similarity between mashups based on the HIN network and leveraged the user-based CF to recommend services more relevant to user requirements.

Nowadays, the hybrid approach integrating multiple models or various types of information to make recommendations has become popular. A common practice is to add side information into MF-based models to improve their performance. For example, considering that some contextual factors influence developers' selection behaviors, Botangen *et al.* [9] derived the relevance scores from two contextual factors: geographical location and textual description. They utilized the relevance scores in an MF-based recommendation model. Similarly, Nguyen *et al.* [10] mined the relationship between services regarding their functional similarities and leveraged this feature to regularize an MF-based model with an attention mechanism. Cao *et al.* [20] proposed a content- and network-based Web API recommendation method by developing a two-level topic model and a co-invocation-based CF algorithm.

Inspired by some well-known models in click-through rate (CTR) prediction, a few new approaches applied deep neural network (DNN) or factorization machine (FM) [11] to service recommendation. For example, Xiong *et al.* [12] combined CF and content-based recommendation models with a DNN. Chen *et al.* [13] presented a neural CF recommender model that learns user preference on services based on their matching degree on explicitly-declared attribute preference and implicit preference mined from historical invocations. By levering the Wikipedia corpus, Cao *et al.* [14] enriched mashups and services in content and then fused features, including the extracted content, similar services or mashups, popularity, and the co-occurrence of services, into an FM to capture their second-order interaction. Furthermore, Cao *et al.* [15] introduced an attention mechanism into an FM and weighed each feature when learning their interactions. Yao *et al.* [21] proposed a probabilistic MF approach with implicit correlation regularization to improve recommendation quality and diversity. The explicit text similarity and implicit correlations between APIs, including the similarity or the complementary relationship between APIs, are integrated into an MF model. Kang *et al.* [22] presented a hybrid FM model integrating a DNN and attention mechanisms, which can capture non-linear feature interactions and their different importance. Yan *et al.* [23] proposed a deep learning-based service recommendation method named CACN (short for collaborative attention convolutional network), which can learn the bilateral information for service recommendation. However, these existing approaches are applicable for mashup development in the single-round scenario, which neglects developer feedback in the development process.

### B. Multi-round service recommendation

In services computing, multi-round interactive design first appeared in service composition. Some researchers adopted the strategy of step-by-step recommendations in their service composition platforms. For example, Zhao *et al.* [24] designed a platform named HyperService, which can search and recommend a set of relevant services for end users according to input keywords and navigation contexts. In particular, service relations were detected and leveraged in the dynamical search process. Since social networks often affect developer selection, Maaradji *et al.* [25] proposed a framework that can acquire knowledge from a social network and incorporate the generated knowledge with user profiles to make recommendations for service discovery. Liu *et al.* [26] applied a sequential pattern mining algorithm to discover frequent service composition patterns. When the proposed approach recommends a candidate sequence, it leverages current user selection and considers both the frequency and the logical order of internal components to facilitate mashup development. Gu *et al*. [27] proposed a composite semantics-based service bundle recommendation model to cover the functional requirements of the target mashup as thoroughly as possible.

These approaches extract service composition patterns from the history of service compositions or the neighborhood in social networks. They leverage such knowledge to calculate the probability of a service that will be selected next according to a user's current selection. Due to the regularity of patterns, the recommendation result obtained after user selection is relatively fixed. Moreover, such a recommendation process often neglects the target users' current requirements.

Besides, user preference evolves with changes in user needs and service functions or quality. A few studies then improved recommendation performance by tracking dynamic preference sequences and predicting user preference in the future. For example, Zhang *et al.* [28] extracted dynamic user preference from time slice data by the time-series latent Dirichlet allocation (LDA) model and generated a service list based on the latest user preference and QoS. Kwapong *et al.* [29] composed a user's invocation preference at a timestamp as a combination of non-functional attributes such as QoS and functional features extracted from the Web services description language (WSDL) file of the invoked service. They modeled user preference sequences and predicted the target user's latest preference for the following recommendation by a long short-term memory (LSTM) network.

However, these methods only focus on the evolution pattern of user selection behavior or preference while neglecting that the next service to recommend, together with selected services, should meet the target mashup's requirements. Xie *et al.* [30] discussed a specific recommendation scenario with selected services in the mashup development process. When developers select one or more services, the user-based CF is leveraged to re-calculate the mashup similarity, find more accurate neighbor mashups, and update their recommendations. Although the CF-based method leverages the interaction between the target mashup and candidate services, it cannot explicitly characterize selected services' critical impacts on developers' selection behaviors.

Inspired by the recent work on user interest representation in CTR prediction (e.g., [31] and [32]), DINRec [33] employs an attention mechanism to weigh selected services according to their relevance to the candidate service. Unfortunately, DINRec is designed to develop mashups with some selected services. In other words, DINRec does not apply to building a new mashup. Besides, it has the following two drawbacks. First, DINRec's prediction model must be retrained after a developer selects a component service. Second, the prediction model takes only discrete features as input, such as the tag attribute of mashups and services. Because DINRec does not consider the textual description of services and mashups, it cannot fully characterize the functional interactions (or called content interactions) between mashups and services.

*C. Multi-round recommendations in recommender systems*

Although multi-round recommendation has emerged as a hot topic in recommender systems in recent years, its application in service recommendation is still rare. Generally speaking, the work in this area can be classified into multi-round interactive and multi-round conversational recommendations, according to whether explicit conversations are involved.

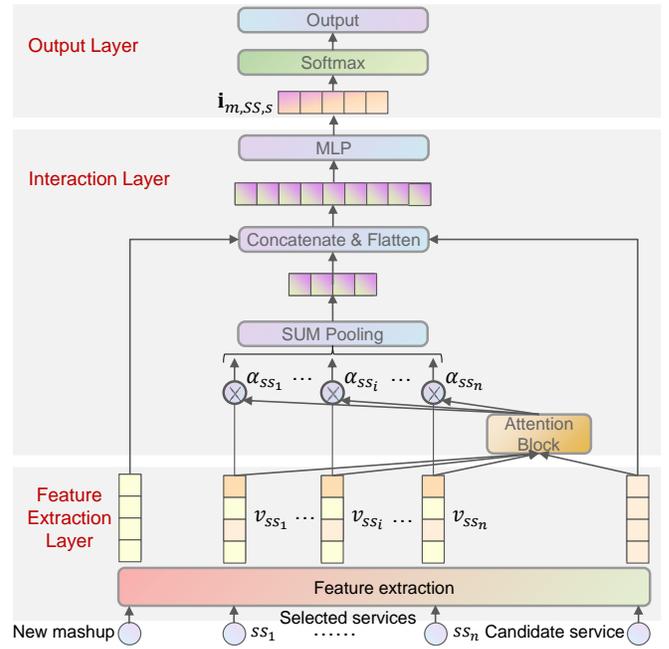

Fig. 2. The architecture of DLISR.

In the multi-round interactive recommendation scenario, the recommender systems capture dynamic user preferences and make recommendations over time [34]-[36]. The recommender waits for user feedback after recommending an item in each round, leading to more accurate follow-up recommendations. In the multi-round conversational recommendation scenario, the recommender asks questions about user preference or makes recommendations repeatedly to provide recommendations with fewer turns of conversation [37]-[39]. Generally speaking, the multi-armed bandit (MAB) and reinforcement learning (RL) are two primary ways to implement multi-round recommendations in recommender systems [38].

However, the multi-round recommendation scenarios differ from those discussed in this study in the following aspects.

First, the multi-round service recommendation aims to find a group of component services (rather than a single item) that can collaboratively satisfy developer requirements.

Second, the RL and MAB methods adopted in the multi-round recommendation scenarios aim to learn uncertain user preferences or intentions by tentative conversations or recommendations. In contrast, user (or developer) requirements in mashup development are usually deterministic and specified.

Third, the multi-round recommendation scenarios require additional information, such as a negative reward and trial-and-error conversations during the recommendation process. However, developers (or users) only provide feedback on whether a recommended service was selected in this study's scenario. An unselected service is often not negative feedback and is still not considered at the current stage of mashup development. Therefore, it is difficult for the MAB and RL methods to be directly used in the multi-round service recommendation scenario.

## III. PROBLEM STATEMENT

A service repository is a collection of mashups and services, represented as $R = M \cup S$, where $M$ is a set of mashups and $S$ is a set of services. Mashups and services have their respective attributes. We leverage the content information of mashups and services (such as textual descriptions and tags) and the provider information of services to make recommendations in this study. An invocation indicates the call relationship between a mashup and its component service. The history of invocations between $M$ and $S$ is regarded as implicit feedback. We represent the invocation history as an invocation matrix $\mathbf{MS}$, where the cell value in the $m$-th row and the $s$-th column is defined as

$$r_{m,s} = \begin{cases} 1, & \text{if } s \text{ is a component service of mashup } m \\ 0, & \text{otherwise} \end{cases} \quad (1)$$

Then, we present the problem to be solved in this study. Suppose a developer (or user) plans to build a new mashup $m$ and has provided keyword-based requirements, referred to as $MReq$. Given a set of selected component services $SS$, how can an SRS recommend an appropriate service bundle to the developer (or user) accurately? Note that $SS$ is empty when the developer starts building $m$.

## IV. AN INTERACTION-ORIENTED FRAMEWORK FOR MULTI-ROUND SERVICE BUNDLE RECOMMENDATION

In the multi-round service recommendation scenario, there exist multiple interactions among keyword-based developer requirements $MReq$, selected services, and the next service to recommend. On the one hand, the candidate and selected services may be replaceable or complementary. On the other hand, they need to work together to satisfy $MReq$. Such interactions help developers determine whether they will select a candidate service as a component service of the target mashup. Therefore, we propose a framework named DLISR to learn these interactions and predict the probability of a developer selecting a candidate service. Then, we briefly introduce the DLISR framework and detail it layer by layer.

### A. Overview of DLISR

Fig. 2 shows the DLISR framework's architecture. It consists of three layers, i.e., a feature extraction layer, an interaction layer, and an output layer.

The feature extraction layer obtains feature representations of a new mashup, each selected service, and the candidate service by taking the original input information. The interaction layer learns the correlation weight between each selected service and the candidate service based on an attention block. After calculating selected services' representations, it employs a multilayer perceptron (MLP) to learn interactions among the target mashup, the candidate service, and selected services. By leveraging the learned interactions, the output layer predicts the possibility that a developer will select the candidate service as a component service in the next round.

When training the DLISR framework offline, we need to construct a training dataset based on all existing mashups and services in a given repository. The training samples are built by simulating how developers gradually select component services for mashup construction. Therefore, the framework trained offline can make online service recommendations efficiently for developing new mashups.

### B. Details of Each Layer in DLISR

#### 1) Feature extraction layer

This layer takes as input the history of invocations between mashups and services or one type of information, such as the content information of a new mashup $m$, a set of selected services $SS$, and a candidate service $s$. It then outputs the corresponding feature representation of the same type to learn about interactions among $m$, $SS$, and $s$. Note that we obtain a feature representation of a mashup or a single service in this layer. We provide two feature extractors for the content information and the invocation history to implement the DLISR framework. For more details, please refer to Section V.

#### 2) Interaction layer

The goal of this layer is to learn about interactions among $m$, $SS$, and $s$. Although there are a few ways to model such interactions, we employ MLPs in this study because they can, in theory, fit arbitrary functions well [40]. Usually, we can employ multiple MLPs to simultaneously learn the interactions among $m$, $s$, and each selected service, and then integrate them with another MLP. However, this straightforward way will consume too many computing resources. Alternatively, we first compress representations of all selected services into a fixed-length vector by utilizing the attention mechanism. As a result, this layer takes only an MLP to learn about interactions among $m$, $s$, and $SS$.

##### a) Attention-based feature integration

The simplest way to comprehensively represent selected services is to concatenate their representations directly. However, this strategy does not fit our task here. On the one hand, the concatenation will increase the final representation's length and reduce the framework's efficiency, especially with many selected services. On the other hand, either the number of selected services or the length of concatenation results is changeable in this study's scenario. Instead, our framework's fully-connected networks used to learn about interactions can only handle fixed-length input.

Another possible solution is to perform average pooling or sum pooling on the representation of each selected service. The pooling method still has its disadvantages, despite generating a fixed-length representation. Let us continue to analyze the case mentioned above, where a developer builds a mashup to help tourists prepare for their trips. Suppose the developer has selected several services, including a hotel reservation service, a weather forecast service, and an online map service. Considering that tourists usually prepay online after booking a hotel, the hotel reservation service will have a more significant impact than the other two services on selecting an electronic payment service as the next component service. Therefore, each selected service has a different impact on selecting the next component service, suggesting that a different weight should be assigned to each service's representation. However, the pooling method weights them equally.

We design an attention-based method to integrate each selected service's representation in this layer. It pays attention to those selected services more relevant to the next recommended service and filters out unnecessary ones. Note that the relevance can be measured by the similarity or complementarity (i.e., different functions complement each other) of service functionalities.

We use the weighted sum of each selected service's features to represent the set of all selected services.

$$\boldsymbol{v}_{SS} = \sum_{s_i \in SS} w_{is} \boldsymbol{v}_i, \quad s.t. \sum_{i \in SS} w_{is} = 1, \quad (2)$$

where $\boldsymbol{v}_i$ is the vector representation of each selected service $s_i$ and $w_{is}$ is the weight of $s_i$. The physical meaning of $w_{is}$ is the correlation degree between $s_i$ and $s$, i.e., the contribution of $s_i$ to the target developer's selection on $s$. So, $w_{is}$ is jointly determined by features of $s_i$ and $s$.

The similarity between two vectors can be measured by their element-wise multiplication or their element-wise subtraction for the difference between them. Given vector representations $\boldsymbol{v}_i$ and $\boldsymbol{v}_s$ of $s_i$ and $s$, the results of these two operations can be used as prior knowledge to help model the correlation between $s_i$ and $s$. We concatenate with features of $s_i$ and $s$ and input the concatenation result into an MLP to learn the correlation between $s_i$ and $s$ automatically. The MLP then outputs a scalar score. The above process can be described as

$$\alpha_{is} = \text{MLP}(\boldsymbol{v}_i \oplus \boldsymbol{v}_s \oplus (\boldsymbol{v}_i \otimes \boldsymbol{v}_s) \oplus (\boldsymbol{v}_i \ominus \boldsymbol{v}_s)), \quad (3)$$

where $\otimes$ is the element-wise multiplication, $\ominus$ is the element-wise subtraction, $\oplus$ is the concatenation operation, and MLP($\cdot$) represents all the default operations within an MLP.

We finally input the score into a softmax function to calculate the final weight $w_{is}$

$$w_{is} = \frac{\exp(\alpha_{is})}{\sum_{s_j \in SS} \exp(\alpha_{js})}. \quad (4)$$

Unlike the average pooling operation, we calculate each selected service's weight adaptively according to its correlation with the candidate service. In this layer, selected services contribute different weights to the overall representation of $\boldsymbol{v}_{SS}$. This representation varies with different candidate services. Our attention-based method can obtain a more adaptable representation than the average pooling operation.

*b)    MLP-based interaction learning*

Up to now, we have obtained representations of $m$, $SS$, and $s$, i.e., $\boldsymbol{v}_m$, $\boldsymbol{v}_{SS}$, and $\boldsymbol{v}_s$, respectively. Because there are no local or sequential patterns in the concatenated representations, convolutional neural networks (CNNs) and recurrent neural networks (RNNs) are not suitable for this interaction learning task [41]. We concatenate these representations and then utilize an MLP to capture interactions among $m$, $SS$, and $s$. Moreover, we select the parametric rectified linear unit (PReLU) as the activation function since it can improve model fitting with nearly zero extra computational cost and little overfitting risk [42]. The learning process can be written as

$$\boldsymbol{i}_{m,SS,s} = \text{MLP}(\boldsymbol{v}_m \oplus \boldsymbol{v}_{SS} \oplus \boldsymbol{v}_s), \quad (5)$$

where $\boldsymbol{i}_{m,SS,s}$ is the learned interaction vector.

*3)    Output layer*

We finally feed the learned interaction vector into a softmax function, whose output ($\hat{r}$) represents the probability of $m$ selecting $s$ as a component service. The process can be written as

$$\hat{r} = \text{softmax}(\boldsymbol{W}^T \boldsymbol{i}_{m,SS,s} + b), \quad (6)$$

where $\boldsymbol{W}$ is a weight matrix and $b$ is the bias parameter.

## V. IMPLEMENTATION MODELS

Given a new mashup $m$, a set of selected services $SS$, and the candidate service $s$, the DLISR framework learns about their interactions and predicts the probability of $m$ selecting $s$ as a component service. We can construct various implementation models by extracting features and learning their corresponding interactions. We will introduce two implementation models: a function-interaction-based service recommendation (FISR) model and a neighbor-interaction-based service recommendation (NISR) model. Also, we present a hybrid model named HISR (short for hybrid-interaction-based service recommendation) that integrates these two models. Therefore, the proposed DLISR framework has three model variants. FISR and NISR apply the DLISR framework to the content information and the history of invocations between mashups and services. In contrast, HISR takes advantage of the content information and the invocation history in the DLISR framework. The ultimate purpose of comparing these model variants is to analyze the impacts of different interactions between mashups and services.

### A. FISR Model

When calculating the selection probability of a component service, developers will first consider whether the service's functionality, together with those of selected services, can satisfy their requirements for the target mashup. Therefore, we design a functional interaction model that captures their interactions from the functional perspective.

A service's functionality derived from its content often has two forms: word sequence (service description) and separate words (tags). The same is true for developer requirements described in natural languages. To obtain the representation of a mashup or a service in functionality, we use two deep-learning-based techniques to process these two forms of content information, respectively, and then concatenate their extracted features.

To apply deep-learning-based feature extraction, we need to preprocess the content information of services and mashups with word embeddings. More specifically, we convert all terms into sparse binary vectors with one-hot encoding, input these vectors into an embedding layer, and map each term into a dense vector or an embedding. If necessary, we would truncate or pad a piece of the content information and then stack the corresponding embeddings of its terms, transforming the content information into a matrix $\boldsymbol{E}$ with a fixed size. The process can be written as

$$\boldsymbol{E} = [\boldsymbol{e}_{t_1}, \boldsymbol{e}_{t_2}, \ldots, \boldsymbol{e}_{t_i}, \ldots, \boldsymbol{e}_{t_L}]^T, \quad (7)$$

where $L$ is the length of the content information, $t_i$ is the $i$-th term in the content information, and $\boldsymbol{e}_{t_i}$ is the word embedding of $t_i$.

We apply the *text_inception* network used in our previous work [43] to the content information represented in a word sequence for feature extraction. This method parallels stacked convolution layers to extract local patterns in a word sequence and then employs the global average pooling (GAP) operation to emphasize important features. Finally, an MLP performs a non-linear transformation on the extracted features. The process to get the representation of a word sequence, $\boldsymbol{v}_{seq}$, can be simply written as

$$\boldsymbol{v}_{seq} = text\_inception(\boldsymbol{E}). \quad (8)$$

There is no order between terms for the content information represented in the form of a separate word set, which makes *text_inception* inapplicable. Instead, we retrieve and average the embedding of each term to get a fixed-length representation of the separate word set, $\boldsymbol{v}_{set}$, described as

$$\boldsymbol{v}_{set} = \text{average}[\boldsymbol{e}_{T_1}, \ldots, \boldsymbol{e}_{T_i}, \ldots, \boldsymbol{e}_{T_M}], \quad (9)$$

where $\boldsymbol{e}_{T_i}$ is the embedding of the $i$-th term in the set, and $M$ is the size of the set.

For each mashup or service, we concatenate the two features extracted from its content information, $\boldsymbol{v}_{seq}$ and $\boldsymbol{v}_{set}$, and obtain a representation that characterizes its functionality.

$$\boldsymbol{v} = \boldsymbol{v}_{seq} \oplus \boldsymbol{v}_{set}. \quad (10)$$

In this way, we can transform the content information of developer requirements, the candidate service, and each selected service into real-valued feature vectors. We then feed these features into the DLISR framework's interaction layer and obtain a vector ($\boldsymbol{ci}_{m,SS,s}$) for their functional interactions. Finally, we input this vector into the output layer to predict a score for the candidate service.

*B. NISR Model*

Besides the content information, the history of invocations between mashups and services benefits good recommendations. Inspired by the user-based CF approach, we design a neighbor-interaction-based model for learning about interactions among $m$, $SS$, and $s$, according to the historical invocation information of mashups similar to the target mashup (i.e., neighbor mashups, denoted as $NM$).

The challenge of applying the DLISR framework to the historical invocation information is effectively obtaining the representations of $m$, $s$, and $SS$. For this purpose, we employ node2vec [44], a typical graph embedding method, to process the graph transformed from the historical invocation matrix $\boldsymbol{MS}$. Compared with those traditional MF-based approaches, such as probabilistic MF [18] and singular value decomposition, node2vec can capture a more non-linear relationship between a mashup and a service.

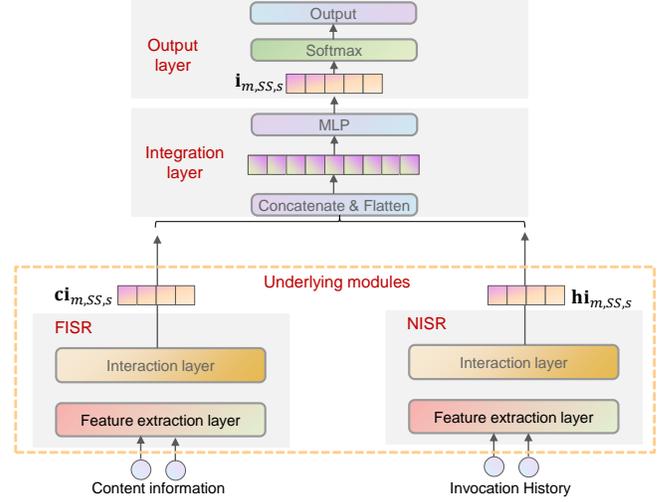

Fig. 3. The architecture of HISR.

In the recommendation scenario of this study, new mashups are built and completed online. The above MF-based methods cannot update their models with newly-selected services in time, nor can they accurately represent the target mashup. Instead, our solution is to search out neighbor mashups for the target mashup and then calculate its representation using neighbor mashups' representations and the similarities between the target mashup and its neighbors. Therefore, the key to this model lies in calculating the similarity between the target mashup $m$ and an existing neighbor mashup $nm_i$.

To this end, we adopt a similar method proposed in [29] to calculate the similarity. We first construct a HIN network to organize mashups and services. Then, we use a meta-path-based method to calculate six types of similarities between two mashups: sharing the same topics, labeled by the same tags, invoking the same service, and invoking similar services with the same topic, tag, or provider. Finally, the weighted sum of these six similarities is calculated as an overall similarity between two mashups. We detail the above method in the Appendix. The similarity weights are set to the pre-trained values in [29]. The overall similarity $sim_{m,nm_i}$ between $m$ and $nm_i$ can be expressed as

$$sim_{m,nm_i} = \sum_{p=1}^{6} w_p * sim_p(m, nm_i), \quad (11)$$

where $sim_p(m, nm_i)$ is the $p$-th meta-path-based similarity and $w_p$ is its corresponding weight.

According to the similarities between the target mashup and existing mashups, we select $K$ mashups most similar to the target mashup as $NM$ and then obtain a weighted representation $\boldsymbol{v}_m$ of $m$.

$$\boldsymbol{v}_m = \sum_{nm_i \in NM} sim_{m,nm_i} \cdot \boldsymbol{v}_{nm_i}, \quad (12)$$

where $\boldsymbol{v}_{nm_i}$ is the representation of $nm_i$ obtained by node2vec.

In addition to the content information of two mashups, this similarity calculation method also considers their component services' information. After a developer selects a component service, the method will improve its similarity measurement and better represent the target mashup to build. Besides, the method that calculates mashup similarities based on meta-paths is efficient. We can also adopt pruning strategies (e.g., reducing the number of candidate mashups) to improve the efficiency of searching for neighbor mashups in a large-scale service repository.

We have now obtained the representations of $m$, each service in $SS$, and $s$ in the same feature space based on the historical invocation information. Then, we can feed them into the DLISR framework's interaction layer and compress their interactions in this space into a dense vector, $\boldsymbol{hi}_{m,SS,s}$. The vector is fed into the output layer to make a prediction.

*C. HISR Model*

The two models learn two forms of interactions among $m$, $SS$, and $s$, according to the content information and invocation history, respectively. This subsection proposes a hybrid model named HISR that integrates these two models to leverage multiple interactions.

The HISR model (shown in Fig. 3) consists of two underlying modules, an integration layer, and an output layer. An underlying module inputs a specific type of information and learns a unique type of interactions among $m$, $SS$, and $s$. More specifically, an underlying module could be viewed as removing the output layer from a separate implementation model (i.e., FISR and NISR) of the DLISR framework. That is to say, each of the two underlying modules has only a feature extraction layer and an interaction layer. Then, the integration layer incorporates all the interactions learned from different underlying modules with an MLP. The process can be expressed as

$$\boldsymbol{i}_{m,SS,s} = \text{MLP}(\boldsymbol{ci}_{m,SS,s} \oplus \boldsymbol{hi}_{m,SS,s}). \quad (13)$$

Finally, the HISR model's output layer makes a prediction based on the learned interactions. It is also worth mentioning that this hybrid model is easily extensible. Suppose a new kind of information is available. In that case, we can add another underlying module to learn a new type of interaction and then integrate it into the original hybrid model to improve model performance further.

*D. Offline Model Training*

The proposed models predict the probability of mashup $m$ selecting service $s$ as the next service to invoke when it has selected a set of component services $SS$. A sample in our study is denoted as $(m, SS, s)$. The predicted value approximates one for a positive sample and zero for a negative one. So, the likelihood function of our models is

$$P(Y^+, Y^-|\Theta) = \prod_{(m,SS,s) \in Y^+} \hat{r}_{(m,SS,s)} \times \prod_{(m,SS,s) \in Y^-} (1 - \hat{r}_{(m,SS,s)}), \quad (14)$$

where $\Theta$ is the model parameter set, $\hat{r}_{(m,SS,s)}$ is the predicted probability for a sample $(m, SS, s)$, and $Y^+$ and $Y^-$ represent positive samples and negative samples, respectively.

The loss function to be minimized is defined as

$$J = -\log P(Y^+, Y^-|\Theta) = -\sum_{(m,SS,s) \in \{Y^+ \cup Y^-\}} [(r_{(m,SS,s)} \log \hat{r}_{(m,SS,s)}) + (1 - r_{(m,SS,s)}) \log(1 - \log \hat{r}_{(m,SS,s)})], \quad (15)$$

where $r_{(m,SS,s)}$ is the actual label of a sample $(m, SS, s)$.

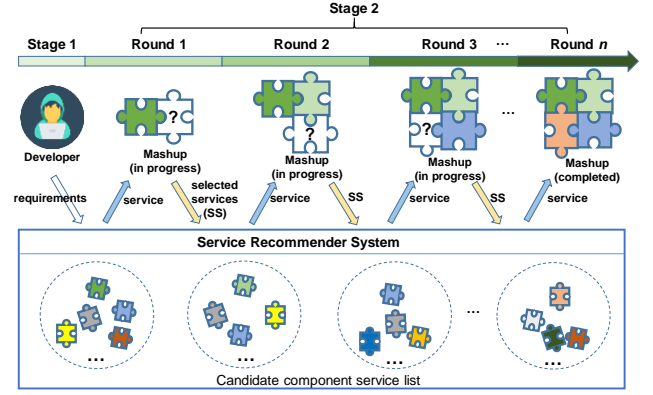

Fig. 4. Scenario setting of experiments.

*Algorithm* 1 presents the model training process of the DLISR framework for the FISR and NISR models. Lines 3-6 show a forward-propagation process to make a prediction. In the third line, we obtain representations of $m$ and $s$ by a feature extraction layer. In the FISR model, we get the representations of mashups and services using Eq. (10). In the NISR model, we obtain the representations of existing mashups and services by node2vec and then calculate the representation of the target mashup $m$ using Eq. (12). In the fourth line, we adopt an attention-based method to get an overall representation of all selected services. In the fifth line, we get the interactions among $m$, $SS$, and $s$. In the sixth line, this algorithm outputs a probability of $m$ selecting $s$ as a component service. Finally, in the seventh line, this algorithm performs back-propagation and employs the Adam algorithm [45] to update all model parameters based on the loss function defined in Eq. (15).

| **Algorithm 1**. Training algorithm of FISR or NISR |
|---|
| **Input**: number of epochs $p$ and sample set $Y$ |
| **Output**: parameter set $\Theta$ |
| 1.  **for** epoch = 1, …, $p$ **do** |
| 2.    **for** each sample composed of mashup $m$, service $s$, and selected services $SS$ in $Y$ **do** |
| 3.      Compute $\boldsymbol{v}_m$ and $\boldsymbol{v}_s$; |
| 4.      Compute $\boldsymbol{v}_{SS}$ using Eq. (2); |
| 5.      Compute $\boldsymbol{i}_{m,SS,s}$ using Eq. (5); |
| 6.      Compute $\hat{r}$ using Eq. (6); |
| 7.      Update $\Theta$ to minimize $J$ in Eq. (15) with the Adam; |
| 8.    **end for** |
| 9.  **end for** |
| 10. **return** $\Theta$; |

Because the HISR model is hierarchical, updating its model parameters may fall into slow convergence. So, we design a transfer-learning-based training algorithm for the hybrid model (see *Algorithm* **2**). This algorithm trains the FISR and NISR models and initializes each underlying module in the hybrid model according to the corresponding pre-trained parameters in FISR and NISR. Next, it freezes the two underlying modules and trains the integration and output layers. Finally, HISR's learnable parameters are unfrozen, and the whole model is fine-tuned.

**Algorithm 2**. Training algorithm of HISR

**Input**: sample set $Y$ and number of epochs $p$
**Output**: parameter set $\Theta$
1. Train FISR and NISR using **Algorithm 1** and get their parameter sets, $\Theta_{FISR}$ and $\Theta_{NISR}$;
2. Initialize HISR's two underlying modules with $\Theta_{FISR}$ and $\Theta_{NISR}$, and then freeze them;
3. Train learnable parameters in the integration layer and the output layer with the input obtained by Eq. (13);
4. Update all the model parameters by fine-tuning the hybrid model;
5. **return** $\Theta$;

## VI. EXPERIMENTAL SETUPS AND RESULT ANALYSIS

### A. Experimental Setups

Fig. 4 illustrates the scenario setting of our experiments. The multi-round service recommendation scenario has two stages. Stage 1 deals with the cold-start problem when a developer initiates building a new mashup. In Stage 2, we consider multi-round cases where the developer has selected one or more component services. This study's experiments were conducted on a workstation with Intel Core 8 Xeon @3.50 GHz, GeForce GTX 2080, and 32 GB memory. The source code implemented based on Keras[2] is available on GitHub[3].

*1) Dataset and samples*

We collected 22,813 APIs and 6,464 mashups from ProgrammableWeb, the largest online Web service registry, on June 30, 2022. We removed all mashups and services without content information from the original dataset. Also, we filtered out any mashups that contained only one component service. The final experimental dataset contains 2,143 mashups, 879 Web services (APIs), and 6,598 mashup-service invocations.

Each experiment sample designed for the multi-round scenario consists of a mashup to build, one or more selected services, and one Web service for testing. The sampling process is defined as follows. Suppose that a new mashup $m$ consists of a few component services $S_m = \{s_1, s_2, ..., s_n\}$, a sample can be described as $(m, CS_i, s_i)$, where $CS_i \in \mathcal{P}(S_m)\setminus\{S_m\}$, i.e., $CS$ belongs to the power set of $S_m$ excluding $S_m$, and $s_i$ is a randomly sampled candidate service. For a positive sample, $s_i \in S_m \setminus CS_i$, while for a negative one, $s_i \in S \setminus S_m$. The difference between a positive sample and a negative one is whether $s_i$ belongs to $S_m$ or not. The size of negative samples is twelve times that of positive samples in our training sample set. Considering that the number of a mashup's component services seldom exceeds four in our experimental dataset, we chose one, two, and three as the number of selected services when building the sample set.

In this study, we evaluated the performance of models using five-fold cross-validation. Mashups in our experimental dataset were divided into five folds, where one fold for testing and the others for training each time.

*2) Evaluation Metrics*

We used the following metrics to evaluate recommendation results and averaged the five-folds' metric values as the final evaluation result.

Three commonly-used metrics, *Precision*, *Recall*, and *F1-measure* at the top $N$ services in a ranking list, are defined as

$$P@N = \frac{1}{|M|} \sum_{m \in M} \frac{|\text{rec}(m) \cap \text{act}(m)|}{|\text{rec}(m)|}, \quad (16)$$

$$R@N = \frac{1}{|M|} \sum_{m \in M} \frac{|\text{rec}(m) \cap \text{act}(m)|}{|\text{act}(m)|}, \quad (17)$$

$$F1@N = \frac{1}{|M|} \sum_{m \in M} 2 \frac{|\text{rec}(m) \cap \text{act}(m)|}{|\text{rec}(m)| + |\text{act}(m)|}, \quad (18)$$

where $M$ is a set of mashups in the test set. For mashup $m$, $\text{rec}(m)$ and $\text{act}(m)$ represent a list of recommended services and a collection of its actual component services, respectively.

*Mean average precision* (*MAP*) at the top $N$ services in a ranking list is defined as

$$MAP@N = \frac{1}{|M|} \sum_{m \in M} \frac{1}{N_m} \sum_{i=1}^{N} (\frac{N_i}{i} \times I(i)), \quad (19)$$

where $I(i)$ represents whether the $i$-th recommended service hits an actual component service of mashup $m$, $N_i$ is the number of hits in the top $i$ positions of the ranking list, and $N_m$ is the number of component services of $m$.

*Normalized discounted cumulative gain* (*NDCG*) at the top $N$ services in a ranking list is defined as

$$NDCG@N = \frac{1}{|M|} \sum_{m \in M} \frac{1}{Q_m} \sum_{i=1}^{N} \frac{2^{I(i)} - 1}{\log_2(1+i)}, \quad (20)$$

where $Q_m$ denotes the ideal maximum DCG score of a recommendation result for mashup $m$.

*3) Baseline Approaches*

We compared several competitive methods to demonstrate our model's effectiveness, covering the content-based, CF-based, and hybrid approaches.
- WVSM (short for weighted vector space model) [46]. This content-based method predicts a candidate service's probability over a mashup according to the WVSM-based similarity between their content information.
- BPR-MF (short for Bayesian personalized ranking for matrix factorization) [47]. This method trains MF-based service recommendation models with a pairwise ranking loss of the Bayesian personalized ranking algorithm. Since

---

[2] https://keras.io

[3] https://github.com/ssea-lab/DLISR

10Table 1. Performance comparison of different approaches in two different stages.

| Method | P@10 | | R@10 | | F1@10 | | NDCG@10 | | MAP@10 | |
|---|---|---|---|---|---|---|---|---|---|---|
| | Stage 1 | Stage 2 | Stage 1 | Stage 2 | Stage 1 | Stage 2 | Stage 1 | Stage 2 | Stage 1 | Stage 2 |
| WVSM | 0.0483 | 0.0304 | 0.1859 | 0.1756 | 0.0747 | 0.0487 | 0.1359 | 0.1099 | 0.0855 | 0.0768 |
| BPR-MF | - | 0.0427 | - | 0.2444 | - | 0.0680 | - | 0.1571 | - | 0.1117 |
| PNCF | 0.0827 | 0.0641 | 0.3143 | 0.3005 | 0.1302 | 0.1003 | 0.3032 | 0.2867 | 0.2271 | 0.2229 |
| SFTN | 0.1199 | 0.0754 | 0.4368 | 0.4139 | 0.1817 | 0.1190 | 0.3721 | 0.2898 | 0.2710 | 0.2196 |
| ISRec | 0.1713 | 0.1002 | 0.6329 | 0.5242 | 0.2611 | 0.1559 | 0.5892 | 0.4102 | 0.4869 | 0.3363 |
| PaSRec | 0.1666 | 0.1025 | 0.6207 | 0.5373 | 0.2546 | 0.1599 | 0.5913 | 0.4303 | 0.4952 | 0.3549 |
| DINRec | 0.1637 | 0.1034 | 0.6026 | 0.5398 | 0.2492 | 0.1612 | 0.5542 | 0.4176 | 0.4493 | 0.3363 |
| FISR | 0.1647 | 0.1023 | 0.6095 | 0.5292 | 0.2513 | 0.1585 | 0.5639 | 0.4130 | 0.4595 | 0.3337 |
| NISR | 0.1669 | 0.1004 | 0.6134 | 0.5238 | 0.2539 | 0.1561 | 0.5829 | 0.4119 | 0.4819 | 0.3346 |
| HISR | **0.1770** | **0.1070** | **0.6619** | **0.5576** | **0.2706** | **0.1667** | **0.6092** | **0.4395** | **0.5019** | **0.3586** |

BPR-MF is a model-based CF approach, it cannot work well in the multi-round service recommendation scenario. Due to the lack of selected component services, it fails to work in the first stage. To enable BPR-MF to work in the second stage, we update it manually whenever developers select a new component service.
- PNCF (short for preference-based neural collaborative filtering service recommendation) [13]. The framework compresses the user and item features in an embedding layer and then uses an MLP to model their interactions. Since its feature extraction component cannot extract text features, we applied the hierarchical Dirichlet process (HDP) in the feature extraction.
- SFTN (short for recommendation through service factors and top-K neighbors) [5]. This method first calculates two probabilities of a mashup invoking a service in the next round according to the content similarity and the historical interaction between neighbor mashups and the service. According to Bayes' theorem, it then multiplies the probabilities as a final rating.
- PaSRec [29]. PaSRec constructs a HIN network and then measures an overall similarity between two mashups based on their meta-paths. Finally, it adopts a user-based CF strategy to predict based on similarity. Besides, this approach designs a pairwise loss function and applies BPR to model optimization.
- ISRec [30]. ISRec improves PaSRec with the measure of content similarity using word embedding. Moreover, it speeds up the search for neighbor mashups by clustering existing mashups offline.
- DINRec [31], [33]. This method applies a deep interest network (DIN) in CTR prediction to Web service recommendations. It exploits available features of the target mashup, selected and candidate services, and then learns their interaction in a well-designed network. We enabled DINRec to utilize the same information and feature extractors as our models for a fair comparison.

*4) Parameter Settings*

The parameters of *text_inception* designed for textual feature extraction in FISR were the same as those in [43]. For node2vec in NISR, the length of node embeddings, walk length, return probability $p$, and the in-out probability $q$ were set to 25, 10, 0.25, and 4, respectively. The pre-trained weights of six meta-path-based similarities were set to 0.14, 0.14, 0.27, 0.15, 0.15, and 0.15. For the MLPs in the attention block, the unit numbers of two hidden layers were set to 80 and 40, respectively. For the MLPs in the interaction layer of FISR or NISR, the unit numbers of two hidden layers were set to 100 and 50, respectively. For the MLPs in the integration layer of HISR, the unit numbers of three hidden layers were set to 128, 64, and 32, respectively. The learning rate was set to 0.0003 for the Adam algorithm [45]. We retained the default parameter settings mentioned in the original references for all the baseline approaches.

*B. Performance comparison of different approaches*

Table 1 presents the performance comparison of different approaches in the first and second stages. As mentioned above, we introduce selected services into samples to evaluate the model performance in the second stage. Stage 2 has three rounds in our experiment, where the numbers of selected services were set to one, two, and three. The average values of the three rounds were taken as the metric values in this stage. Numbers shown in bold font indicate the best results across all the approaches.

FISR and WVSM are two approaches using only the content information. The former performed much better than the latter regarding the five evaluation metrics. On the one hand, FISR uses a CNN-based feature extractor to obtain high-quality and task-specific features. On the other hand, FISR considers selected services' functionalities when learning the interactions between mashups and services.

BPR-MF and NISR utilize only the historical invocation information. The recommendation result of BPR-MF is much worse than that of NISR, suggesting that the model-based CF does not work well in this specific scenario. BPR-MF needs to update its model according to the target mashup's selected services to obtain the embedding of a mashup built online. However, the number of selected services is usually too small for BPR-MF to obtain a high-quality embedding of the target mashup, which leads to poor recommendation results.

SFTN and PNCF leverage historical invocation and content information to make recommendations. Similar to BPR-MF, they use different collaborative filtering techniques to deal with historical invocations. They improved the performance of BPR-MF (see Table 1) because of the usage of the content information. However, they performed worse than other hybrid approaches.

Table 2. Performance comparison of different methods in all the rounds of Stage 2.

| Round No. | Model | P@10 | Gain | R@10 | Gain | F1@10 | Gain | NDCG@10 | Gain | MAP@10 | Gain |
|---|---|---|---|---|---|---|---|---|---|---|---|
| 1 | ISRec | 0.1066 | -0.28% | 0.6325 | -1.33% | 0.1717 | -0.35% | 0.525 | -2.19% | 0.4555 | -3.45% |
| | PaSRec | 0.1058 | 0.47% | 0.6303 | -0.98% | 0.1708 | 0.18% | 0.5248 | -2.15% | 0.4553 | -3.40% |
| | DINRec | 0.1023 | 3.91% | 0.6005 | 3.93% | 0.1647 | 3.89% | 0.4859 | 5.68% | 0.4121 | 6.72% |
| | HISR | 0.1063 | - | 0.6241 | - | 0.1711 | - | 0.5135 | - | 0.4398 | - |
| 2 | ISRec | 0.0985 | 6.80% | 0.5065 | 7.72% | 0.1527 | 7.20% | 0.3862 | 8.88% | 0.3108 | 8.91% |
| | PaSRec | 0.1000 | 5.20% | 0.5168 | 5.57% | 0.1555 | 5.27% | 0.406 | 3.57% | 0.3287 | 2.98% |
| | DINRec | 0.1018 | 3.34% | 0.5282 | 3.29% | 0.1585 | 3.28% | 0.4017 | 4.68% | 0.3199 | 5.81% |
| | HISR | 0.1052 | - | 0.5456 | - | 0.1637 | - | 0.4205 | - | 0.3385 | - |
| 3 | ISRec | 0.0956 | 14.54% | 0.4335 | 16.06% | 0.1434 | 15.20% | 0.3193 | 20.42% | 0.2425 | 22.68% |
| | PaSRec | 0.1018 | 7.56% | 0.4649 | 8.22% | 0.1533 | 7.76% | 0.3601 | 6.78% | 0.2806 | 6.02% |
| | DINRec | 0.1060 | 3.30% | 0.4906 | 2.55% | 0.1604 | 2.99% | 0.3652 | 5.28% | 0.2768 | 7.48% |
| | HISR | 0.1095 | - | 0.5031 | - | 0.1652 | - | 0.3845 | - | 0.2975 | - |

DINRec, PaSRec, and ISRec have their respective advantages regarding the given indicators. For example, PaSRec outstripped DINRec and ISRec in terms of NDCG and MAP. Compared with them, HISR achieved better results across all the evaluation metrics in the two stages, indicating its effectiveness over the baselines.

As shown in Table 1, FISR and NISR outperformed WVSM and BPR-MF, respectively. Meanwhile, HISR performed the best in the hybrid models. Therefore, we believe that the models following the DLISR framework perform better than the baselines using the same information, indicating the effectiveness of the DLISR framework.

Furthermore, we compared HISR and three competitive hybrid approaches, namely DINRec, ISRec, and PaSRec. Table 2 presents the indicator values in the three rounds of Stage 2. The symbol "Gain" denotes the percentage of the performance improvement of HISR relative to the target baseline.

When the developer selected only one component service, HISR did not perform better than ISRec and PaSRec regarding MAP, NDCG, and Recall. The reason is that the solely-selected service contributed little helpful information or even noise to the learning process of HISR. When the number of selected services increased to two, HISR achieved the best results across all the given indicators. Moreover, as the number of selected services increased to three, our hybrid approach got advantages over the three baselines. The reasons are analyzed as follows.

PaSRec and ISRec estimate one candidate service's scalar score over a new mashup according to the history of invocations between neighbor mashups and the service. However, the two baselines do not use the interactions between selected services and the mashup (or the service). Instead, HISR learns the interactions among the target mashup, selected services, and the candidate service by an elaborate interaction layer equipped with an MLP and an attention block.

DINRec was proposed based on a DIN model designed by Zhou *et al.* [31] for e-commerce recommendation. Considering that each item's features are usually sparse and homogeneous, the original DIN enforces all kinds of item features to interact with each other sufficiently to learn latent interaction patterns. In this experiment, features of mashups and services were extracted from two feature spaces (i.e., the content information and historical invocations). DINRec shows a limited ability to learn the interactions between features from two independent and heterogeneous feature spaces. In contrast, HISR learns internal interactions in the two feature spaces and fuses them with an MLP, thus increasing the learning performance.

We used the Wilcoxon signed-rank test [48], a commonly-used non-parametric test, to analyze the difference between HISR and DINRec, the best baseline method, regarding the $F1@10$ and $NDCG@10$ metrics. We repeated ten independent runs of five-fold cross-validation for the two methods in Stage 2 of the multi-round recommendation scenario. The statistical significance tests for 50 pairs of prediction results indicated statistically significant differences in the two metrics between ours and DINRec at a 99% confidence level ($p < 0.01$).

### C. Integration strategies of selected services

DLISR predicts the probability of a mashup with selected component services invoking a service based on their complex interactions. To learn such interactions more efficiently, we need to integrate each selected service's representation and comprehensively represent them. An attention mechanism is utilized in our framework to accomplish this task. To compare the impacts of integration strategies on recommendation performance, we replaced the attention mechanism with three other strategies and generated three framework variants.

- DLISR-Average. This framework performs the average pooling strategy on the representation of each selected service and assigns equal weight to them.
- DLISR-Concatenation. This framework directly concatenates the representation of each selected service. We truncate or pad the set of selected services to get a final representation with a fixed size.
- DLISR-None. This framework disables selected services during the model learning process.

We evaluated the performance of DLISR and its three different variants in all three rounds of Stage 2. Table 3 shows the comparison result and highlights the "Gain" of our framework relative to the target baseline.

DLISR-None performed the worst in most rounds, indicating that selected services did play an essential role in the multi-round service recommendation. When developers selected only one service, there was no apparent difference in performance among DLISR, DLISR-Concatenation, and DLISR-Average. When developers selected two and more services, our attention-based framework performed better than DLISR-Average and DLISR-Concatenation. The above result indicates that DLISR can pay more attention to those selected services related to the current prediction and obtain adaptive representations specific to different candidate services. Our approach aims to help develop large-size mashups, but the number of selected component services in our experimental dataset is relatively small. Therefore, the performance improvement of adopting the attention mechanism is not as significant as expected. This assertion can be verified by the observation that the percentage of performance improvement gradually increases as the number of selected services increases in most rounds.

Table 3. Performance comparison between DLISR and its variants.

| Model | Round No. | Content information | | | | | | Historical invocation information | | | | | |
|---|---|---|---|---|---|---|---|---|---|---|---|---|---|
| | | F1@10 | Gain | NDCG@10 | Gain | MAP@10 | Gain | F1@10 | Gain | NDCG@10 | Gain | MAP@10 | Gain |
| DLISR | 1 | 0.1645 | - | 0.4887 | - | 0.4164 | - | 0.1622 | - | 0.4822 | - | 0.4106 | - |
| | 2 | 0.1564 | - | 0.3924 | - | 0.3119 | - | 0.1545 | - | 0.3915 | - | 0.3119 | - |
| | 3 | 0.1546 | - | 0.3578 | - | 0.2728 | - | 0.1516 | - | 0.3618 | - | 0.2812 | - |
| DLISR-None | 1 | 0.1616 | 1.79% | 0.4785 | 2.13% | 0.4055 | 2.69% | 0.1619 | 0.19% | 0.4785 | 0.77% | 0.4165 | -1.42% |
| | 2 | 0.1463 | 6.90% | 0.365 | 7.51% | 0.2873 | 8.56% | 0.1488 | 3.83% | 0.3709 | 5.55% | 0.3097 | 0.71% |
| | 3 | 0.1404 | 10.11% | 0.3095 | 15.61% | 0.2282 | 19.54% | 0.1428 | 6.16% | 0.332 | 8.98% | 0.2547 | 10.40% |
| DLISR-Concatenation | 1 | 0.1627 | 1.11% | 0.4846 | 0.85% | 0.4143 | 0.51% | 0.1608 | 0.87% | 0.4744 | 1.64% | 0.4015 | 2.27% |
| | 2 | 0.1518 | 3.03% | 0.3808 | 3.05% | 0.3032 | 2.87% | 0.1476 | 4.67% | 0.3699 | 5.84% | 0.2936 | 6.23% |
| | 3 | 0.1489 | 3.83% | 0.3433 | 4.22% | 0.2629 | 3.77% | 0.1445 | 4.91% | 0.3439 | 5.21% | 0.2678 | 5.00% |
| DLISR-Average | 1 | 0.1635 | 0.61% | 0.4822 | 1.35% | 0.4091 | 1.78% | 0.1615 | 0.43% | 0.4764 | 1.22% | 0.4022 | 2.09% |
| | 2 | 0.1524 | 2.62% | 0.3835 | 2.32% | 0.3048 | 2.33% | 0.152 | 1.64% | 0.3859 | 1.45% | 0.3106 | 0.42% |
| | 3 | 0.1483 | 4.25% | 0.3387 | 5.64% | 0.2591 | 5.29% | 0.1471 | 3.06% | 0.3496 | 3.49% | 0.2734 | 2.85% |

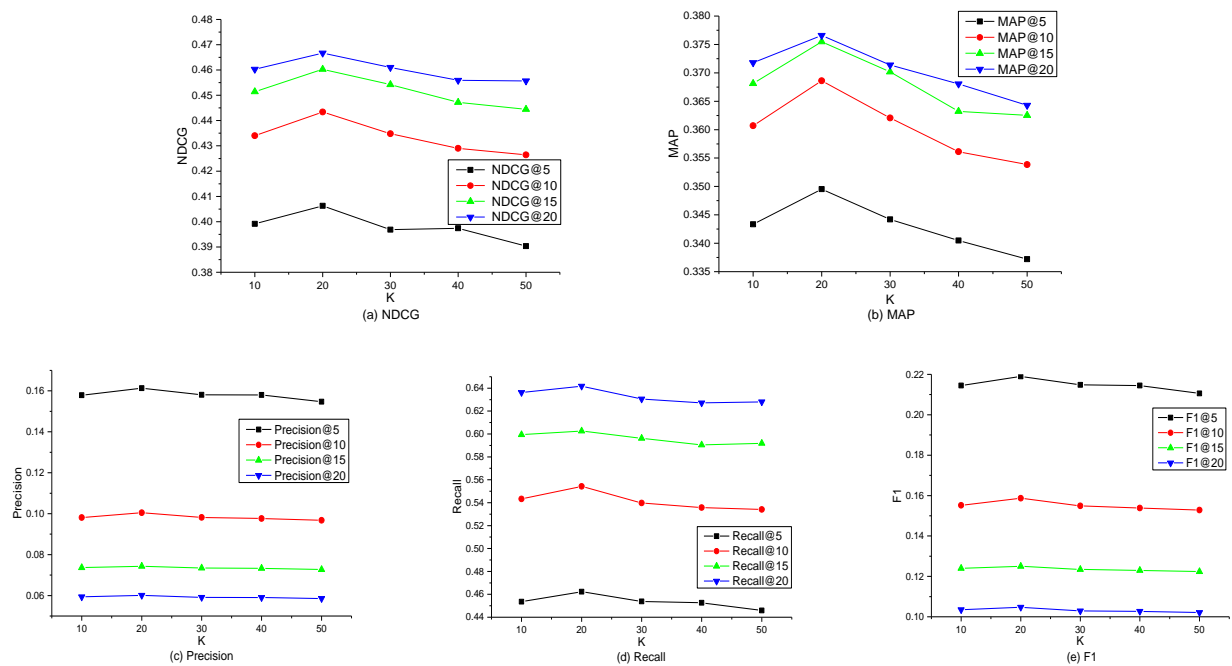

Fig. 5. Recommendation performance with different $K$

### D. Impact of the Size of Neighbor Mashups

Due to the cold-start problem, it is hard to directly model the interaction between a new mashup and a candidate service. We approximate it with the interaction between the new mashup's neighbors and the candidate service. Therefore, the number of neighbor mashups, $K$, is a critical factor in applying the DLISR framework to the historical invocation information. To study this parameter's effect on recommendation performance, we adjusted $K$ from 10 to 50 with step 10.

As shown in Fig. 5, all the indicators increase as the value of $K$ increases from 10 to 20, suggesting that our approach can exploit more beneficial information from neighbor interactions. However, the opposite is true when $K$ exceeds 20. Perhaps the introduction of noisy data brings harmful interference to the interaction learning at this stage. Therefore, we set $K$ to 20 in our experiments.

*E. Discussion*

According to the experimental results mentioned above, we can draw the following conclusions:
- HISR outperforms other competitive baseline approaches in the first and second stages of the multi-round service bundle recommendation scenario for iterative mashup development.
- Selected services do affect the selection of candidate services. Furthermore, when integrating representations of selected services, the attention mechanism performs better than the commonly-used concatenation and average-pooling strategies in most rounds.

The proposed approach is designed for mashup development. As mentioned before, a mashup is a kind of Web application that can provide certain functionalities by composing one or more services. Therefore, our proposed approach can be used in developing other Web-based applications, as long as they contain any existing Web services or APIs as their components.

However, our approach still has some limitations. The first limitation is that the experiments were performed on a single source dataset. ProgrammableWeb is the sole service repository containing publically available interactions between mashups and services. We performed experiments on this dataset using five-fold cross-validation to mitigate the evaluation bias. In the future, we will test our proposed approach on more Web service datasets of similar domains to demonstrate its generalizability.

Another limitation is that we did not compare the proposed approach with those classical multi-round recommendation methods. It is unfeasible for us to directly apply MAB and RL, two primary multi-round recommendation methods, to the multi-round service recommendation scenario discussed in this study. Moreover, those multi-round recommendation methods are also inappropriate for the comparison experiment. The main reasons are analyzed as follows. First, the experiment of this paper simulated the state of the first, second, and third round of interactions through the status that a mashup selects one, two, and three component services, respectively. However, this process only considers the selection of correct component services at each round and neglects the selection of incorrect ones. The interaction data in the trial and error process is essential but unavailable for research. In other words, this dataset can only provide what will be the next one, given that some services have been selected. We cannot simulate such data by ourselves to demonstrate the superiority of our approach over MAB and RL methods. Second, the classical multi-round recommendation methods leverage the feedback of incorrect selection at each round to updating their models. Therefore, the current dataset cannot satisfy the training process of classical multi-round recommendation methods, and it is hard for us to choose these multi-round recommendation methods for comparison in this study.

## VII. Conclusion

This study highlights a multi-round service recommendation scenario for iterative mashup development. We propose a deep-learning-based service bundle recommendation framework to address the problems in this scenario. An attention mechanism weighs selected component services when selecting a candidate service. The framework can thus learn the interaction among the target mashup, selected services, and the candidate service. Finally, it predicts the probability of the target mashup invoking a candidate service in the subsequent round of recommendations. According to the framework, we design two separate models for learning such interactions from the perspectives of content and historical invocations. Then, we integrate these two types of interactions in a hybrid model. Experiments indicate that the hybrid model outperforms several state-of-the-art service recommendation approaches.

We plan to improve our approach from the following three aspects in the future. Firstly, frequent local patterns may exist in developers' selection behaviors [49]. For example, after a developer invokes a service, the developer has a high probability of invoking other services related to the selected service. We will explicitly introduce these patterns into our framework and implementation models. Secondly, the multi-round service bundle recommendation is investigated only from the functional perspective. We plan to incorporate QoS into the recommendation process and maximize performance rewards by the RL technology [50]. Thirdly, due to the lack of such experimental data, the DLISR framework does not leverage developers' personalized preferences or long-term interests. Supposing these developers' data are available, we can improve our model by integrating it with an extra score concerning developers' long-term preferences and personal attributes to make a more accurate prediction.

## Acknowledgment

This work was supported by the National Key Research and Development Program of China (No. 2020AAA0107705) and the National Science Foundation of China (Nos. 61972292 and 62032016). Jian Wang is the corresponding author of this paper.

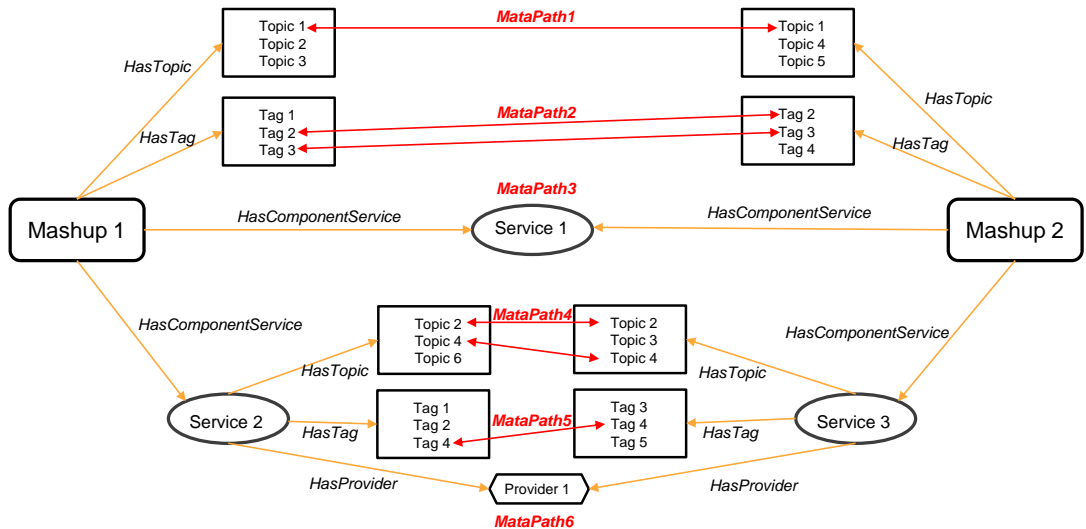

Fig. A1. The structure of the heterogeneous information network.

APPENDIX: HIN-BASED SIMILARITY BETWEEN MASHUPS

We build a heterogeneous information network (HIN) to organize existing entities in a service registry, e.g., mashups and services. Each mashup or service has its content information, which generally falls into two categories: word sequence (service description) and separate words (tags).

We first process the content information in the form of word sequence by LDA and use the top three latent topics obtained as a straightforward and approximate representation of the content information. We use the *hasTopic* edge in the HIN to connect mashups or services and their corresponding latent topics. Meanwhile, we use the *hasTag* edge to connect mashups or services and their content information in the form of separate words. Considering each service has a provider, we use the *hasProvider* edge to connect services and their providers. Since invocations exist between mashups and services, we use the *HasComponentService* edge to connect mashups and their component services. Fig. A1 shows the structure of our HIN, which includes six types of meta-paths to establish connections between two mashups:

- MetaPath1: mashup1-topic-mashup2 means that two mashups share the same latent topic.
- MetaPath2: mashup1-tag-mashup2 means that two mashups have the same tag.
- MetaPath3: mashup1-service-mashup2 means that two mashups invoke the same service.
- MetaPath4: mashup1-service1-topic-service2-mashup2 means that two mashups invoke similar services that share the same topic.
- MetaPath5: mashup1-service1-tag-service2-mashup2 means that two mashups invoke similar services with the same tag.
- MetaPath6: mashup1-service1-provider-service2-mashup2 means that two mashups invoke similar services released by the same provider.

We employ the path-based method [51] to calculate the similarity between mashups. Taking MetaPath1 between mashups $m_1$ and $m_2$ as an example. The similarity based on this path, $sim_p(m_1, m_2)$, can be calculated using the following formula:

$$sim_p(m_1, m_2) = \frac{2 \times |topics(m_1) \cap topics(m_2)|}{|topics(m_1)| + |topics(m_2)|}, \quad (A1)$$

where $topics(m_1)$ and $topics(m_2)$ are the latent topic sets of $m_1$ and $m_2$, respectively.

In this way, we can obtain all the similarities between two mashups based on the six meta-paths and calculate their weighted sum to get an overall similarity between the two mashups.